# Golden ratio autocorrelation function and the exponential decay

Roumen Tsekov

Department of Physical Chemistry, University of Sofia, 1164 Sofia, Bulgaria

An autocorrelation function is obtained on the base of the recurrence relation formalism, whose continued fraction form corresponds to that of golden ratio. It turns out that this GR autocorrelation is known in science and obeys all necessary conditions, in contrast to the exponential autocorrelation function. Using the Kubo approach it is shown how exponential correlations appear in the linear response theory as a result of non-Hermitian relaxation of the system.

The golden ratio (GR) is a magic proportion widely spread in the Nature, architecture, music, etc. From mathematical point of view, it is the positive root of the following quadratic equation $x^2 = x+1$ and equals numerically to $\varphi = (1+\sqrt{5})/2$. The negative root $\phi = (1-\sqrt{5})/2$ of the equation above is known as the GR conjugate and is related to GR via $\varphi + \phi = 1$. Using the relationship $\varphi = 1 + 1/\varphi$, which follows from this quadratic equation, one can present GR in continued fraction form

$$\varphi - 1 = \cfrac{1}{1+(\varphi-1)} = \cfrac{1}{1+\cfrac{1}{1+\cfrac{1}{1+\cdots}}} \qquad (1)$$

The Fibonacci chains are other special numbers related to GR, which can be easily generated via the Binet formula $F_n = (\varphi^n - \phi^n)/(\varphi - \phi)$. The Fibonacci numbers are applied in computational algorithms, coding, random number generators, quasipcrystals, etc. Recently, the Fibonacci chains are used for modeling the momentum autocorrelation function of finite number oscillators [1].

It is well known from the recurrence relation formalism [2-4] that the Laplace image $S(z)$ of an autocorrelation function (ACF) $C(t)$ can be always presented as continued fraction

$$S = \cfrac{1}{z\tau_1 + \cfrac{1}{z\tau_2 + \cdots}} \qquad (2)$$

where $z$ is the Laplace transformation variable. This equation is a slight modification of the original recurrence relation expression, where $\{\tau_k\}$ is a set of characteristic relaxation time constants of the consecutive random forces. Since Eq. (2) resembles Eq. (1), we are tempted to find out the autocorrelation function, which corresponds to GR. Among all possible ACFs with infinite dimensional space the simplest one is hyperspherical [5]. In this case, all the characteristic time constants possess one and the same value, i.e. $\tau \equiv \tau_1 = \tau_2 = \cdots$ and thus Eq. (2) simplifies to

$$S = \frac{1}{z\tau + S} \tag{3}$$

The relationship of this Laplace spectral density to GR is obvious, because $S = \varphi - 1$ at $z\tau = 1$. Moreover, Eq. (3) shows equivalence of the spectral densities of the random process and the corresponding fluctuation force [5], which correlates well to the observed self-similarity in ACFs built by the use of GR [6].

One can analytically solve Eq. (3) and the solution reads

$$S = \sqrt{1 + (z\tau/2)^2} - z\tau/2 = \frac{1}{\sqrt{1 + (z\tau/2)^2} + z\tau/2} \tag{4}$$

This function is appropriate for a Laplace spectrum since it is positive, finite at $z = 0$ and tends to zero at infinity. One can easily check again that $S$ reduces to GR at the characteristic frequency, i.e. $S(z\tau = 1) = -\phi$ and $S(z\tau = -1) = \varphi$. Thus, $-S(z)$ corresponds to the GR conjugate, while GR corresponds to $S(-z)$ and they are related each other via the relation $S(z) = 1/S(-z)$, similar to $-\phi = 1/\varphi$. The inverse Laplace transformation of Eq. (4) is possible and the corresponding GR ACF acquires the analytical form

$$C = J_1(2t/\tau)/t \tag{5}$$

where $J_1(\cdot)$ is the Bessel function of first kind and first order. The GR autocorrelation (5) possesses a finite dispersion $C(0) = 1/\tau$. Its plot in Fig. 1 shows a sequence of anticorrelations and correlations modulated by a long-time tail. The amplitude of $C$ vanishes at large times by a power law $t^{-3/2}$, which is observed in many physical systems and computer simulations. The Fourier spectral density of this ACF is $\mathrm{Re}[S(z=i\omega)]=\sqrt{1-(\omega\tau/2)^2}$ within the frequency range $-2/\tau\leq\omega\leq 2/\tau$ and zero outside. A particular example of the GR ACF is one with zero correlation time $\tau=0$. In this case the spectral density from Eq. (4) is constant $S=1$, which corresponds to a white noise with a Dirac delta ACF $C=2\delta(t)$. The GR autocorrelation (5) is rigorously derived by Rubin for a chain of harmonic oscillators [7]. It is applied as a memory kernel in hydrodynamics [8], Brownian motion of particles [9] and living cells [10], financial markets [11], etc.

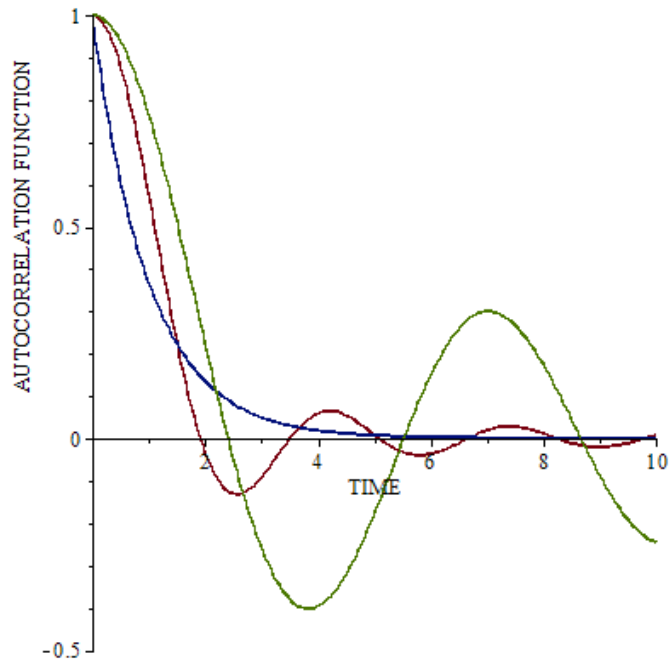

**Fig. 1** The normalized ACFs vs. dimensionless time $t/\tau$:

Exponential $\exp(-t/\tau)$ (blue), GR $J_1(2t/\tau)(\tau/t)$ (red) and $J_0(t/\tau)$ (green)

Let us now consider another process, which is driven by a GR noise. In this case the correlation time of the process $\tau\equiv\tau_1$ differs from the correlation time of the noise $\tau_2=\tau_3=\cdots$. The corresponding Laplace spectral density acquires the form

$$S=\frac{1}{z\tau+\sqrt{1+(z\tau_2/2)^2}-z\tau_2/2} \qquad (6)$$

This expression describes, for instance, the velocity ACF of an impurity in a chain of harmonic oscillators [7, 12]. Equation (6) contains several important particular cases. If the Langevin noise is white ($\tau_2 = 0$) the corresponding ACF $C = \exp(-t/\tau)/\tau$ decays exponentially (see Fig. 1). The GR ACF (5) follows naturally from Eq. (6) at $\tau_2 = \tau$. If the noise correlation time is twice larger than the process ones ($\tau_2 = 2\tau$), the spectral density (6) reduces to $S = 1/\sqrt{1+(z\tau)^2}$ and the corresponding ACF $C = J_0(t/\tau)/\tau$ is also plotted in Fig. 1. When the correlation time of the Langevin force $\tau_2$ is much larger than $\tau$ the spectral density (6) tends to $S = 1/z\tau$ and its constant ACF $C = 1/\tau$ implies a deterministic process. Finally, the Laplace image (4) of the GR ACF corresponds to Eq. (6) also in the unusual case of $\tau = 0$ and a negative $\tau_2 < 0$. The golden ratio is just a number and, for this reason, the GR autocorrelation function provides it only at some special points. The reason we called this ACF the golden ratio ones is the hyperspherical type of its Laplace spectral density (2), which resembles Eq. (1). Moreover, the structure of Eq. (4) looks pretty similar to the golden ratio roots structure. In addition, the GR ACF implies self-similarity between the autocorrelation functions of the considered process and its Langevin force (see Eq. (3)). Perhaps, this is also a cause for the magic character of the golden ratio as well.

From equilibrium statistical mechanics, it is well known that ACF of an arbitrary dynamic variable $V(p,q)$, being a function of the momenta and coordinates of the system particles, can be calculated via the expression

$$C = <V \exp(\hat{L}t)V> = \sum_{k=0}^{\infty} \frac{<V\hat{L}^k V>}{k!} t^k \qquad (7)$$

where $\hat{L} \equiv \{H, \cdot\}$ is the Hermitian Liouville operator of the whole system and the brackets $<\cdot>$ indicate statistical average by the equilibrium distribution density. Since the latter satisfies the stationary Liouville equation, the statistical moments in Eq. (7) can be expressed by integration by parts as follows

$$<V\hat{L}^k V> = (-1)^k <(\hat{L}^k V)V> = (-1)^k <V\hat{L}^k V> \qquad (8)$$

From this relation immediately follows that the moments with odd $k$ are equal to zero. Therefore, the autocorrelation function (7) acquires a more specific form [4]

$$C = \sum_{k=0}^{\infty} \frac{<V\hat{L}^{2k}V>}{2k!} t^{2k} = \sum_{k=0}^{\infty} (-1)^k \frac{<(\hat{L}^k V)^2>}{2k!} t^{2k} \qquad (9)$$

A rigorous property following from Eq. (9) is that the autocorrelation function possesses a maximum at the beginning, i.e. $\partial_t C = 0$ at $t = 0$. Note also the alternation of sign of the terms. Since the Liouville operator is Hermitian it follows that $C(t) = C(-t)$.

Expanding Eq. (5) in series of time confirms that the GR ACF obeys Eq. (9) as well. This is not the case, however, of the widely used exponential ACF $C = \exp(-t/\tau)/\tau$, which does not seem to be a proper one [13]. On the other hand, according to the Doob theorem [14] the exponential ACF is compulsory for stationary Gaussian Markov processes. Since the Gaussian character follows from the central limit theorem, it seems that Markov processes contradict mechanics in general. Perhaps, this is a new manifestation of the Loschmid paradox proving again that thermodynamics is not simply a mechanics of many particle systems. The exponential ACFs are usually measured in the linear response theory. To clarify the problem let us consider the following Liouville equation

$$\partial_t \rho = \hat{L}\rho + \hat{O}\rho \qquad (10)$$

which describes the evolution of the probability density $\rho$ towards the equilibrium canonical distribution $\rho_{eq} = \exp(-\beta H)/Z$. Since the relaxation operator $\hat{O}$ is not Hermitian, it is clear that thermodynamic relaxation is irreversible. In what follows we are going to trail the Kubo approach. Note that in the original Kubo paper [15] the operator $\hat{O}$ is missing and it is not clear then how the canonical Gibbs distribution is established. Supposing, at the initial moment an external force is applied to the system, being at equilibrium. Then the Hamilton function of the system will change to $H + \Delta H$, which will reflect in a new distribution density $\rho = \rho_{eq} + \Delta\rho$. Introducing these expressions in Eq. (10) leads to

$$\partial_t \Delta\rho = (\hat{L} + \hat{O})\Delta\rho + \Delta\hat{L}\rho_{eq} \qquad (11)$$

where the nonlinear term $\Delta\hat{L}\Delta\rho$ is neglected in the frames of a linear analysis and the properties of the equilibrium Gibbs distribution $\partial_t\rho_{eq}=0$, $\hat{L}\rho_{eq}=0$ and $\hat{O}\rho_{eq}=0$ are employed.

The integration of Eq. (11), under the initial condition $\Delta\rho=0$ at $t=0$, is straightforward and the solution reads

$$\Delta\rho = \int_0^t \exp[(\hat{L}+\hat{O})(t-s)]\Delta\hat{L}\rho_{eq}ds \tag{12}$$

Using the properties of the canonical Gibbs distribution the term $\Delta\hat{L}\rho_{eq}$ can be elaborated in the form $\beta\rho_{eq}\hat{L}\Delta H$ and introducing it in Eq. (12) yields

$$\Delta\rho = \beta\rho_{eq}\int_0^t \exp[(\hat{L}+\hat{O})(t-s)]\hat{L}\Delta H ds \tag{13}$$

Let us suppose now that the external force $F(t)$ acts on a single particle. In this case the contribution to the Hamilton function will be $\Delta H = -XF(t)$, where $X$ is the coordinate of the target particle. Using the expression for the particle velocity $V=-\hat{L}X$ additionally, Eq. (13) changes to

$$\Delta\rho = \beta\rho_{eq}\int_0^t \exp[(\hat{L}+\hat{O})(t-s)]VF(s)ds \tag{14}$$

One can calculate by Eq. (14) the average change in the velocity of the target particle under the action of the external force

$$<\Delta V> = \int V\Delta\rho dpdr = \beta\int_0^t <V\exp[(\hat{L}+\hat{O})(t-s)]V> F(s)ds \tag{15}$$

According to the Kubo theory [15] one is able to determine the equilibrium fluctuations in the system by measuring of the linear response. Looking at Eq. (15) one can recognize the measured velocity ACF of the target particle in the form

$$C = <V \exp[(\hat{L}+\hat{O})t]V> \tag{16}$$

In the standard Kubo approach the relaxation operator is zero and Eq. (16) coincides with Eq. (7). Since the relaxation operator $\hat{O}$ is not Hermitian an analog of Eq. (8) does not exist and hence the ACF (16) cannot be presented in the form (9). Hence, the exponential relaxation is not forbidden. Indeed, if we consider the simplest BGK relaxation operator $\hat{O}\rho = (\rho_{eq} - \rho)/\tau$ [16] the analog of Eq. (16) will read

$$C = <V \exp(\hat{L}t)V> \exp(-t/\tau) \tag{17}$$

As is seen ACF (7) is exponentially modulated here. If the considered probability density $\rho$ is a single-particle one, then $\hat{L} = -V\partial_X$ and the term $<V \exp(\hat{L}t)V> = <V^2>$ is constant. Thus Eq. (17) reduces to an exponential ACF. The same result follows from the Markovian Fokker-Planck relaxation operator, while a more complicate measured ACF (16) is expected when $\hat{O}$ is the non-linear Boltzmann collision operator.

The disagreement between mechanics and thermodynamics is also evident from the Second Law of thermodynamics. Introducing the Shannon entropy [17] via $S \equiv -k_B \int \rho \ln \rho \, dpdq$, the entropy production equals to

$$\partial_t S = -k_B \int \partial_t \rho \ln \rho \, dpdq = -k_B \int \hat{O}\rho \ln \rho \, dpdq \tag{18}$$

where the last expression is obtained by using Eq. (10). Hence, if the relaxation operator $\hat{O}$ is zero, the entropy of an isolated system is constant anytime, which contradicts the Second Law. To overtake this shortcoming, some authors argue the non-equilibrium entropy definition, while others propose coarse graining or time-sliding procedures. In the case of a Fokker-Planck relaxation operator $\hat{O} = \partial_p(mk_BT\partial_p + p)/\tau$, Eq. (18) simplifies to

$$\partial_t S = k_B(-mk_BT\int \rho \partial_p^2 \ln\rho \, dpdq - 1)/\tau \tag{19}$$

It is intersecting that the production of Shannon information is proportional to the Fisher information [18]. Since $\partial_t S \geq 0$ according to the Second Law, it follows from Eq. (19) that the momentum Fisher information $F_p \equiv -\int \rho \partial_p^2 \ln\rho \, dpdq$ decreases in time. At equilibrium $F_p = \beta/m$, which leads naturally to the canonical Gibbs distribution. The same situation holds in the coordinate space as well, where the free Brownian motion obeys the classical diffusion equation

$$\partial_t c = D\partial_q^2 c \tag{20}$$

Thus, the production of configuration entropy $S = -k_B \int c \ln c \, dq$ is proportional to the configuration Fisher information $F_q = -\int c \partial_q^2 \ln c \, dq$, i.e. $\partial_t S = Dk_B F_q$. Since the solution of Eq. (20) is a normal distribution density, the Fisher information $F_q = 1/2Dt$ also decreases in time.

Another general problem is if the relaxation processes affect the equilibrium distribution. In this case the latter will be a solution of the stationary Eq. (10) $(\hat{L}+\hat{O})\rho_{eq} = 0$ but not of separate equations $\hat{L}\rho_{eq} = \hat{O}\rho_{eq} = 0$. In fact, the canonical Gibbs distribution is valid only for the case of weak coupling of the system to the environment [19, 20]. To see how the environment affects the system distribution let us consider first that the whole system is isolated. Than $\hat{L}\rho_{eq} = 0$ and the equilibrium solution is the well-known microcanonical Gibbs distribution

$$\rho_{eq} = \delta(E-H)/\Omega \tag{21}$$

Let us divide now the system into a subsystem S and a bath B. The Hamilton function of the whole system reads $H = H_S + U_{SB} + H_B$, where $U_{SB}$ is a potential describing the subsystem-bath interaction. Integrating the microcanonical Gibbs distribution (21) along the bath particles momenta and coordinates yields the distribution of the subsystem particles

$$\rho_S = \int \rho_{eq} dp_B dq_B \sim \int (1 - \frac{H_S + U_{SB} + U_B}{E})^{\frac{3}{2}N} dq_B \qquad (22)$$

The last expression here is derived via explicit integration over the bath particles momenta. In the thermodynamic limit the number of bath particles $N$ is infinitely large and hence the full energy is a linear function of $N$, i.e. $E = 3N/2\beta$. Thus, Eq. (22) acquires the form

$$\rho_S \sim \exp(-\beta H_S) \int \exp[-\beta(U_B + U_{SB})] dq_B = \exp[-\beta(H_S + F_{BS})]/Z \qquad (23)$$

One can easily recognize in $F_{BS}$ the conditional free energy of the bath at a fixed configuration of the subsystem particles. It describes the average effect of the subsystem-bath interaction and, in contrast to the usual potentials, it is temperature dependent. Note that the generalized canonical Gibbs distribution (23) does not obey the subsystem Liouville equation, i.e. $\hat{L}_S \rho_S \neq 0$. The Cooper pairs in superconductors are very popular manifestations of interaction mediated by the bath particles [21]. Since the interactions among the particles depend only on the distance between them, $F_{BS}$ is constant, when the subsystem consists of a single particle. Hence, the Maxwell-Boltzmann distribution is not affected by the subsystem-bath interaction. Imagine now we are in a position to switch off the bath. Then the subsystem becomes isolated and its evolution is governed by the Liouville equation $\partial_t \rho_S = \hat{L}_S \rho_S$ coupled with an initial condition from Eq. (23). If the initial condition is the classical canonical Gibbs distribution, the latter would be the solution of this Liouville equation at any time. Hence, the subsystem-bath interaction term $F_{BS}$ is only responsible for evolution to the right microcanonical Gibbs distribution $\rho_{eq} = \delta(E_S - H_S)/\Omega_S$ at equilibrium (see the Appendix). An alternative point of view on the problem how to construct the potential function $H_S + F_{BS}$ is proposed by Yuan and Ao [22, 23], which elucidates also the dissipative character of the bath-subsystem interaction.

**Appendix**

Hereafter the microcanonical Gibbs distribution (21) is derived from classical mechanics. According to the Poincare theorem, a mechanical system always returns very near to its initial state. Such quasi-periodic systems are conveniently described via action $I$ and angle $\varphi$ variables [24]. Since the Hamiltonian function of quasi-periodic systems $H(I)$ depends only on the action, the integration of the corresponding Hamilton equations $\dot{\varphi} = \partial_I H$ and $\dot{I} = -\partial_\varphi H$ is straightforward. The action $I$ remains constant equal to its initial value $I_0$, while the angle increases linearly in time, $\varphi = \varphi_0 + (\partial_I H)_{I_0} t$. Therefore, the microscopic probability density acquires the form

$$\rho = \delta(I - I_0)\delta(\varphi - \varphi_0 - (\partial_I H)_{I_0} t) \tag{A1}$$

As is seen from Eq. (A1), $\rho$ is continuous function of time, which fluctuates permanently. Hence, there is no equilibrium distribution in mechanics. The equilibrium thermodynamic state could be attributed to the most frequently observed microscopic state. Therefore, the equilibrium thermodynamic distribution is a time average of microscopic probability density

$$\rho_{eq} \equiv \lim_{T \to \infty} \frac{1}{T} \int_0^T \rho(I, \varphi, t) dt \tag{A2}$$

Introducing Eq. (A1) in Eq. (A2) and employing properties of the Dirac delta function leads to

$$\rho_{eq} = \delta((\partial_I H)_{I_0}(I - I_0)) \lim_{T \to \infty} \frac{1}{T} \int_0^T \delta(t - (\partial_H I)_{I_0}(\varphi - \varphi_0)) dt = \delta(E - H)/\Omega \tag{A3}$$

where the energy of the system is defined by $E = H(I_0)$. As is seen, this is the microcanonical Gibbs distribution (21), where all non-additive integrals of motion vanished and the system energy $E$ remains the only one active. It follows from the definition (A2) that the average value of any quantity by ensemble coincides with the average value on time. Therefore, the ergodic theorem is always fulfilled for quasi-periodic systems. It is known that the GR autocorrelation function corresponds to an ergodic process [25].

An interesting consequence from Eq. (A3) is that the time averaging results in some metaphysical correlations. Imagine two non-interacting systems are set together. While the systems are statistically independent in mechanics and $\rho(1\cap 2)=\rho(1)\rho(2)$, it follows from Eq. (A2) a statistical correlation in thermodynamics, where $\rho_{eq}(1\cap 2)\neq \rho_{eq}(1)\rho_{eq}(2)$. This effect explains the positive entropy of mixing in thermodynamics and, perhaps, the KAM theory can through light on statistical interactions between almost non-interacting systems. According to Eq. (A2) the thermodynamic equilibrium state is a superposition of many most frequently observed (most probable) mechanical states. Such a picture corresponds to the Boltzmann view and supports the time-sliding solution of the entropy production problem, discussed before. This is not surprising since any thermodynamic measurement requires a finite time, which is always larger than some resonances in the system. The latter are, in general, very short in many particle systems.

According to the quantum mechanics the most complete description of a quantum system is given in terms of the wave function. For this reason, the classical notion of phase space probability density is replaced by the density matrix operator $\hat{\rho}$. In the case of the Schrödinger equation the density matrix operator obeys the von Neumann equation. Its formal solution in the energy basis acquires the following form

$$\hat{\rho}=\sum\sum \exp[i(E_n-E_k)t/\hbar]|E_k\rangle\langle E_k|\hat{\rho}(0)|E_n\rangle\langle E_n| \qquad (A4)$$

where $\{E_n\}$ are the energy eigenvalues of the system Hamiltonian. As is seen from Eq. (A4) the density matrix possesses non-diagonal elements, while the equilibrium density matrix, following from the quantum statistical physics, is diagonal $\hat{\rho}_{eq}=\sum p_k|E_k\rangle\langle E_k|$ with $p_k$ being the probability for occupation of the state $|E_k\rangle$. The density matrix from Eq. (A4) is a periodic function of time, thus reflecting the Poincare cycles as well. In fact, the evolution never stops and the stationary equilibrium distribution is an idealization, when fluctuations are somehow suppressed. It is obvious that one could eliminate the effect of the persistent fluctuations by averaging in time in the form of Eq. (A2). According to this definition the equilibrium distribution is the most frequently occupied one. Introducing the density matrix operator from Eq. (A4) and performing the integration on time leads straightforward to

$$\hat{\rho}_{eq}=\sum \delta_{E_n E_k}|E_k\rangle\langle E_k|\hat{\rho}(0)|E_n\rangle\langle E_n|=\sum |E_k\rangle\langle E_k|\hat{\rho}(0)|E_k\rangle\langle E_k| \qquad (A5)$$

where the last expression presumes a non-degenerated energy spectrum of the system. Identifying the probability density $p_k = \langle E_k | \hat{\rho}(0) | E_k \rangle = \delta_{EE_k}$ of the microcanonical ensemble, Eq. (A5) reduces to the diagonal expression known from the equilibrium quantum statistical physics. The consideration above shows that decoherence in isolated systems is caused by the quantum evolution itself and the averaging in time leads to mutual cancelation of the non-diagonal fluctuating elements. It is expected that this self-decoherence mechanism takes place in open systems as well, thus assisting decoherence caused by the environment [26].